\documentclass[twocolumn,showpacs,preprintnumbers,amsmath,amssymb]{revtex4}
\usepackage{subfigure}     
\usepackage{graphicx}

\begin{document}

\title{The quest for dark matter with neutrino telescopes\footnote{Prepared for "Neutrino Astronomy- Current status, future prospects", 
 Eds. T. Gaisser \& A. Karle (World Scientific)}}

\author{Carlos P\'erez de los Heros}

\affiliation{Department of Physics and Astronomy,\\
Uppsala University, Uppsala, Sweden.\\
cph@physics.uu.se
}

\begin{abstract}
  There should be not doubt by now that neutrino telescopes are competitive instruments when it comes to searches for dark matter. Their 
large detector volumes collect hundreds of neutrinos per day. They scrutinize the whole sky continuously, being sensitive 
to neutrino signals of all flavours from dark matter annihilations in nearby objects (Sun, Earth, Milky Way Center and Halo) as well 
as from far away galaxies or galaxy clusters, and over a wide energy range. In this review we summarize the analysis techniques and recent 
results on dark matter searches from the neutrino telescopes currently in operation. 

\end{abstract}

\maketitle

\section{Introduction}\label{sec:dm_intro}

The need for a dark matter component in the  universe  is now overwhelming, but so far the indications arise  
from gravitational effects only: rotation curves of galaxies, gravitational lensing in clusters of galaxies or 
structure formation seeded by density fluctuations in the early universe, as derived from cosmic microwave background measurements (see e.g.~\cite{Bergstrom:2000pn}). 
The fact is that dark matter must contribute to the energy budget of the universe approximately five times more than ordinary matter.  
Concrete evidence for any particular type of dark matter is, though, still lacking.  
Searches for dark matter are usually focused on scenarios where the candidates consist of stable relic particles 
whose present density is determined by the thermal history of the early universe. Such an approach is further justified by the fact 
that extensions of the Standard Model predict the existence of particles with the right interaction strength and quantum numbers 
required of a generic candidate for dark matter. The circle thus closes, and our theories of the smallest 
components of matter connect seamlessly with our understanding of the evolution of the universe at grand scales: a beautiful aspect which is hard to ignore. 

In practice the particle dark matter paradigm just needs a stable (or sufficiently long-lived)  massive particle with weak 
interactions, generically called a WIMP (Weakly Interacting Massive Particle). The mass and the couplings of WIMPs with baryonic matter are  
free parameters as far it concerns the astrophysical problem to be solved. These quantities are specified by the underlying particle physics theory 
and need to be determined experimentally.  Good WIMP candidates arise in Supersymmetry~\cite{Nilles:1983ge,Martin:1997ns}: from the neutralino in the Minimal 
Supersymmetric Standard Model (MSSM)~\cite{Djouadi:1998di,Rodriguez:2009cd} to the lightest particle in models with extra dimensions~\cite{Cheng:2002ej,Bertone:2002ms}, 
or models with R-parity violation where an unstable gravitino is the dark matter candidate. 
A feature of gravitino dark matter is that it would leave no signal in direct-detection experiments since the 
cross-section for the interaction between a gravitino and baryonic matter is suppressed by the Planck mass to the fourth power~\cite{Covi:2008jy,Grefe:2011kh}. 

However, the lack of evidence so far for supersymmetry from direct searches, first at LEP~\cite{Kraan:2005vy}, then at the Tevatron~\cite{Toback:2014tea} 
and more recently at the LHC~\cite{Cakir:2015gya}, has restricted the allowed phase space of the theory and raised the  supersymmetric particle mass 
scale to the  $\cal{O}$(TeV) region. There are other flavours of supersymmetry, like the phenomenological MSSM (pMSSM)~\cite{Berger:2008cq} or the Next-to-Minimal 
Supersymmetric Standard Model (NMSSM)~\cite{Ellwanger:1993xa} which can still accommodate WIMPs down to the few GeV region. 

There are extensive reviews in the literature about the particle physics connection of the dark matter 
problem from the theoretical/phenomenological point of view~\cite{Jungman:1995df,Feng:2000zu,Bertone:2004pz,Feng:2010gw,Bergstrom:2012fi}. 
This being a ``Neutrino Astronomy'' issue, we will concentrate on reviewing the latest experimental results from neutrino telescopes.

\section{Indirect dark matter searches with neutrino telescopes}\label{sec:dm_general}

 The riveting possibility of detecting dark matter with neutrino telescopes is based on the fact that it can be 
gravitationally trapped in the deep gravitational wells of heavy objects, like the Sun, the Earth and the halos of galaxies. 
Since dark matter candidates are electrically neutral they can be their own antiparticles, and subsequent pair-wise annihilation into Standard Model 
particles could lead to a detectable neutrino flux. This is a clear signal for a neutrino telescope: it is directional and has a different energy 
spectrum than the known atmospheric neutrino background flux. For such scenario to be viable, the WIMP must have some type of interaction with 
baryonic matter (for the capture in heavy objects to occur) and have a certain level of annihilation cross section (for a neutrino signal to 
be produced). Dark matter candidates might not be their own antiparticles. In this scenario, called 
``asymmetric dark matter''~\cite{Davoudiasl:2012uw,Petraki:2013wwa}, the current dark matter halos would be populated with just dark matter 
(that is, not dark anti-matter) and no annihilation would then be taking place. This picture needs a mechanism to produce only dark-matter at 
the beginning of the universe, in a way similar to baryogenesis, and it would be bad news for indirect searches~\cite{Graesser:2011wi}.

 The actual spectrum of neutrinos detectable at the  Earth depends on the underlying particle physics model used to 
describe the WIMP, through the branching ratios to different final states~\cite{Bottino:1994xp,Cirelli:2005gh,Blennow:2011zz}. 
Neutrinos arise from the annihilation of WIMPs into quarks, charged leptons or gauge bosons. Hadronization or decays of the annihilation 
products will produce a  neutrino flux with an energy dependence determined by the annihilation channel. 
Note that high energetic neutrinos (above a few TeV) produced from the products of 
annihilations of WIMPs inside the Sun will undergo neutral and charged current interactions in the dense solar interior on their way out, and 
the resulting outgoing flux is skewed to lower energies with respect to the original spectrum. That is not the 
case for searches from the center of the Earth or the Galaxy, where the amount of material is not enough to distort the original 
annihilation spectrum.

\section{Current Experiments}\label{sec:dm_expts}

 There are currently three large-scale underwater/ice Cherenkov neutrino telescopes in operation, IceCube at the South Pole~\cite{Achterberg:2006md}, 
ANTARES off the coast of Tulon, France~\cite{Collaboration:2011nsa} and Baikal, in Lake Baikal, Russia,~\cite{Avrorin:2011zza}. These are 
open-volume detectors in the sense that the instrumentation is deployed into a huge volume of naturally occurring water or ice.
Such approach is the most cost-effective to instrument volumes of $\cal{O}($km$^3$). 
At a lower scale, we have Super-Kamiokande~\cite{Abe:2013gga} using 50~kT of ultra-pure water in a vessel located 1.000~m underground in the Mozumi mine in Kamioka. 
The detection method is similar for all these detectors: neutrinos are detected by the Cherenkov light emitted by secondary particles produced in a neutrino 
interaction inside or near the detector. The relative timing of the signals in the photo-multiplier tubes that surround the detector volume allow to reconstruct the 
direction of the original neutrino, and the amount of light deposited is related to the neutrino energy. 
Also underground, but using scintillator instead of water, is the Baksan array~\cite{Alekseev:1998ib}, situated in the North Caucasus at a depth 
of  850 meters of water equivalent. It consists of several planes of scintillators distributed in a four-storey 17~m$\times$17~m$\times$11~m cavern. 
The detector reconstructs upgoing muons measuring the time-of-flight through the detector planes. Baksan is the detector that has been running for a longer time, since 1977.

 The main background of any analysis with neutrino telescopes is the overwhelming flux of muons produced in cosmic-ray interactions in the 
atmosphere, {\it atmospheric muons}.  The same interactions produce a flux of neutrinos, {\it atmospheric neutrinos} which constitute 
an irreducible background to any search for new physics. Atmospheric muons can be filtered out by using the Earth as a filter, at the 
expense of reducing the sky coverage of the instrument.  Full-sky coverage can be regained by defining a ``veto region'' in order to tag 
incoming tracks, most probably an atmospheric muon, and define ``starting events'', which must have been produced by a neutrino interaction inside the 
detector. This comes at the price of reducing the effective volume of the detector and having a somewhat different energy response for 
upgoing and downgoing events. For high enough neutrino energies ($\cal{O}$(10)~TeV), the possibility exists of rejecting atmospheric neutrinos 
when accompanied by a muon produced in the same air shower ~\cite{Schonert:2008is}.

\section{Dark matter searches from the Sun and Earth}\label{sec:dm_sunearth}

 WIMPs that may have accumulated  gravitationally during the lifetime of the solar system in the center of the Sun or Earth, 
can annihilate producing a measurable neutrino 
flux~\cite{Press:1985ug,Krauss:1985aaa,Srednicki:1986vj,Gaisser:1986ha,Ritz:1987mh,Gould:1987ww,Gould:1987ir,Gould:1988eq,Bergstrom:1998xh,Lundberg:2004dn}. 
While any other product of the annihilation will be absorbed, neutrinos will not, and a neutrino telescope can ``look'' 
inside the Sun or the Earth a signal of such annihilations. The strength of the expected neutrino flux depends on several factors, not least 
on the inter-relationship between the capture rate of WIMPs, $\Gamma_{\rm C}$, proportional to the WIMP-nucleon cross section, and the annihilation 
rate, $\Gamma_{\rm A}$, proportional to the velocity averaged WIMP-WIMP annihilation cross section. 
 WIMPs will in general have spin and can then interact with baryonic matter through a spin-dependent and a spin-independent coupling, 
which arise from  axial and scalar terms in the Lagrangian, respectively~\cite{Engel:1992bf}.  Since the Sun is primarily a proton 
target (75\% of H and 24\% of He in mass)~\cite{Grevesse:1998bj} the capture of WIMPs from the halo can be considered to be driven mainly via the 
spin-dependent scattering. Other, heavier, elements constitute less than 2\% of the mass of the Sun, but can still play a role 
when considering spin-independent capture since the spin-independent cross section, $\sigma_{SI}$, is proportional to the squared of the 
atomic mass number. 

For the Earth the situation is rather different. The most abundant isotopes of the main components of the Earth inner core, mantle and 
crust, $^{56}Fe$, $^{28}Si$ and $^{16}O$~\cite{Herndon:80a}, are spin 0 nuclei. Furthermore, the escape velocity at the Earth surface is just 14.8 km/s at 
its center. These values lie at the lower tail of the expected local WIMP velocity distribution, which is assumed to have a mean of the 
order of 220 km/s. Taken at face value, the Earth would appear to be very inefficient in trapping dark matter particles from the 
halo. But the composition of the Earth comes to the rescue, at least for WIMP masses which are resonant with the atomic mass of the 
main components of the Earth~\cite{Gould:1987ir}, which favours the capture of relatively low-mass WIMPs ($m_{\chi} \lesssim$ 50 GeV).

Indeed, the number of WIMP annihilations in the Sun or Earth, N, varies with time 
as $\dot{N}= \Gamma_{\rm C} - \Gamma_{\rm A} N^2/2$.   An additional evaporation term from WIMP-nucleus scattering could be included, 
but it is negligible for WIMP masses above a few GeV~\cite{Krauss:1985aaa,Griest:1986yu}.
Given the age of the Sun (4.5 Gyr), the estimated local dark matter density ($\rho_{\rm{local}}\sim 0.4 GeV/cm^3$) and a weak-scale 
interaction between dark matter and baryons, many models predict that dark matter capture and annihilation in the Sun have reached 
equilibrium. Thus, annihilation is at its maximum possible value, $\Gamma_{\rm A} = \Gamma_{\rm C}/2$.

Experimentally, what a neutrino telescope measures are muons and particle showers from neutral and charged-current neutrino interactions inside or near 
the detector.  We can relate the WIMP annihilation rate $\Gamma_{\rm A}$ in the Earth or Sun and the neutrino flux 
at the detector, $\Phi_{\nu}$, above a given energy threshold $E_{\rm thr}$ as 
\begin{equation} 
\Phi_{\nu}(E_{\nu}) =  \frac{d N_\nu}{d E_\nu \,dA\,dt} = \frac{\Gamma_{\rm A}}{4\pi D^2} \int_{E_{\rm thr}}^{\infty} dE'_{\nu} \, \epsilon(E'_{\nu}; E_{\nu}) \left( \frac{dN_{\nu}}{dE'_{\nu}} \right)
\label{eq:nuflux}
\end{equation}
where $dN_{\nu}/dE'_{\nu}$ is the all-flavour neutrino flux at the center of the source and $\epsilon(E'_{\nu};E_{\nu})$ is a factor that takes 
into account the probability for a neutrino of energy $E'_{\nu}$ to loose energy on its way out of the Sun/Earth and reach the detector with 
an energy $E_{\nu} < E'_{\nu}$. This factor is not needed for the Earth case, but it becomes relevant for the dense 
interior of the Sun. Most of the experimental searches use the muon channel, since it gives better pointing and, in the end, dark matter searches 
from the Sun or Earth are really point-source searches. In that case an additional factor, $P_{\rm osc}^{\rm i, \mu}$, the oscillation probability of 
flavour $i$ to a muon neutrino, needs to be factored in.
\begin{figure*}[t]
\centerline{
{\includegraphics[width=0.5\linewidth,height=0.4\linewidth]{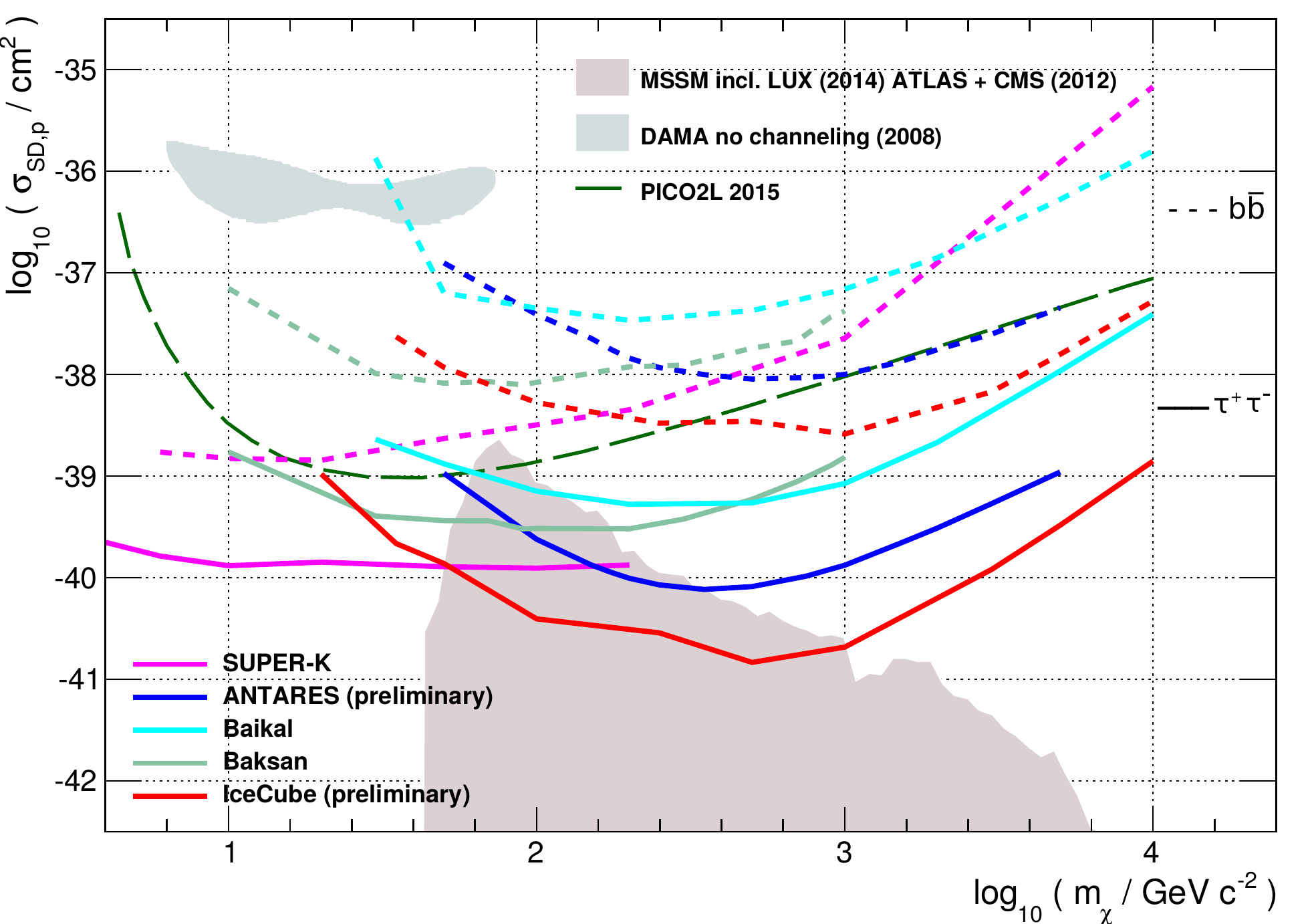}}
{\includegraphics[width=0.5\linewidth,height=0.4\linewidth]{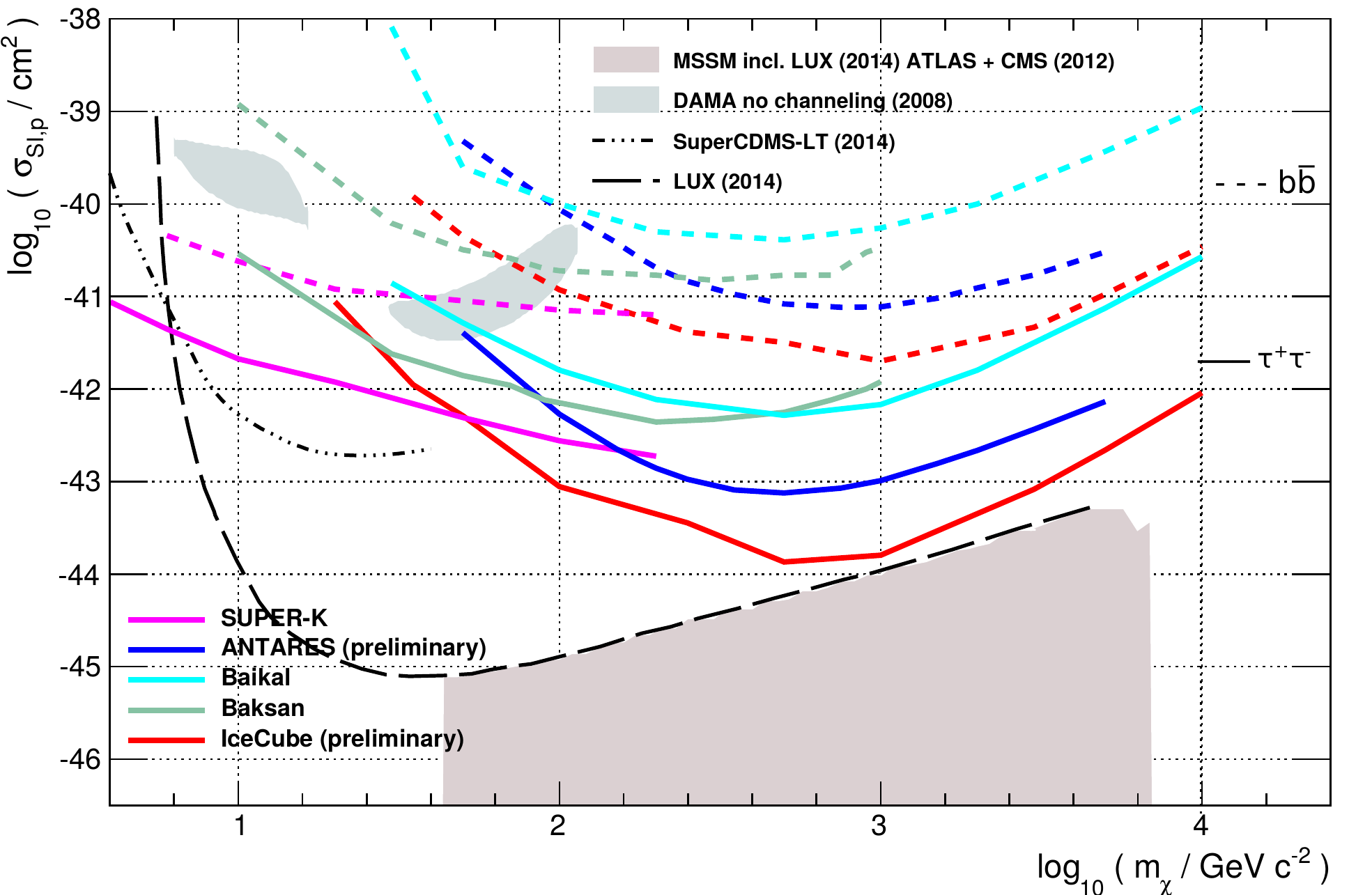}}
}
\caption{
Upper limits at 90\% CL on the spin-dependent (left) and spin-independent (right) WIMP-proton cross section as a function of WIMP 
mass. Limits from IceCube~\cite{Aarsten:2015b}, Super-K~\cite{Choi:2015ara}, ANTARES~\cite{ANTARES:2015a}, Baikal~\cite{Avrorin:2014swy} and 
Baksan~\cite{Boliev:2013ai} are shown. Full lines refer to limits on the $\tau^+\tau^-$ annihilation channel and dashed lines to the $b\bar{b}$ channel.
Direct search results from PICO~\cite{Amole:2015lsj} and LUX~\cite{Akerib:2013tjd}, and tentative signal regions~\cite{Savage:2008er,Aalseth:2011wp,Agnese:2013rvf} 
(gray-shaded areas) are included for comparison.  The brown-shaded region indicates the parameter space from  a 25-parameter MSSM scan~\cite{Silverwood:2012tp}.
}
\label{fig:SunLimits}
\end{figure*}

 In practice, the figure of merit of neutrino telescopes is  the effective area, $A^{\rm eff}_\nu$, the equivalent area with which it would 
detect a neutrino with 100\% efficiency. The effective area is energy dependent and much smaller than the geometrical area of the detector, 
since it includes the neutrino-nucleon cross section and trigger and analysis efficiencies. It can only be calculated with the help of 
detailed detector Monte Carlo simulations and it is always given for a specific analysis and signal type. Equation~\ref{eq:nuflux} can 
be then translated into the more familiar form for the number of events expected at the detector from a neutrino flux $d N_\nu / d E_\nu \,dA\,dt$ 
produced at the source,
\begin{equation}
 N_\nu =  T_{\rm live} \int^{\infty}_{E_{\nu}^{\rm thr}} d E_\nu A^{\rm eff}_\nu(E_\nu) \frac{d N_\nu}{d E_\nu \,dA\,dt}
\label{eq:nevents}
\end{equation}
 where $T_{\rm live}$ is the exposure time of the detector, and now the dependence on the annihilation rate at the source is incorporated in 
the calculation of the neutrino flux.  Under the assumption that the capture rate is fully dominated either by the spin--dependent or spin--independent scattering, 
conservative limits can be extracted on either the spin-dependent or spin-independent  WIMP--proton cross section from the limit 
on $\Gamma_{\rm A}$~\cite{Wikstrom:2009kw}. 
Cross sections are  useful quantities since they allow an easy comparison with the results of direct searches or predictions of a specific 
particle physics model. Such conversion introduces an additional systematic uncertainty in the calculation of the cross sections, due to the element 
composition of the Sun or Earth, the effect of planets on the capture of WIMPS from the halo~\cite{Sivertsson:2012qj} and nuclear form factors used 
in the capture calculations~\cite{Engel:1992bf,Bottino:1999ei,Ellis:2008hf,deAustri:2013saa}. 

\subsection{Current results}
 Searches for dark matter accumulated in the center of the Sun  have been carried out by 
ANTARES~\cite{Adrian-Martinez:2013ayv}, Baikal~\cite{Avrorin:2014swy}, Baksan~\cite{Boliev:2013ai}, Super-K~\cite{Choi:2015ara} and 
IceCube~\cite{Aartsen:2012kia,Aarsten:2015b,Aarsten:2015c}, and a summary of their results is shown in figure~\ref{fig:SunLimits}.
 Since the exact branching ratios of WIMP annihilation into different channels is model-dependent, experiments usually choose two annihilation 
channels which give extreme neutrino spectra to show their results. Annihilation into $b\bar{b}$ is chosen as a representative case producing a soft 
neutrino spectrum, and annihilation into W$^+$W$^-$ or $\tau^+\tau^-$ as a hard spectrum. Assuming a 
100\% branching ratio to each of these channels brackets the expected neutrino spectrum from any more realistic model with branching to more 
channels. The full and dashed curves  in figure~\ref{fig:SunLimits} illustrate this. Since large-volume neutrino telescopes 
are high-energy neutrino detectors, the sensitivity increases by more than an order of magnitude between the ``soft'' and ``hard'' spectra and, in both 
cases, it decreases rapidly with decreasing WIMP mass (softer neutrino spectra). The limits for the spin-dependent cross section are competitive though. 
Direct search experiments do not reach cross section values below 10$^{-39}$ cm$^2$ at their best point, worsening rapidly away from it. IceCube or SuperK 
reach bounds at the $\sim$10$^{-40}$-10$^{-41}$ cm$^2$ level, covering between the two experiments the WIMP mass range from a few GeV to 100~TeV. 

 The picture changes dramatically when we consider spin-independent limits. Here direct-search experiments have the advantage of dedicated spinless targets, 
and the limits from neutrino telescopes lie about three orders of magnitude above the best limit from Lux at a WIMP mass of about 50 GeV. The situation improves slightly 
 for higher masses. But even if the limits of direct experiments worsen rapidly away from the resonance interaction with their 
target nucleus, the limits from neutrino telescopes on the spin-independent cross section lie above over the whole mass range.

\section{Dark matter searches from Galaxies}\label{sec:dm_galaxies}

\subsection{The halo issue}

The accepted scenario for the formation of cosmic structures assumes the formation of regions of increased dark matter  density through 
gravitational collapse from primordial density fluctuations, which in turn attract atomic gas, seeding the formation of galaxies~\cite{Knobel:2012wa}. 
This scenario favours cold (or warm) dark matter over a relativistic species, since in the latter case the formation of large-scale structures 
would have been suppressed and we would not recover the observed universe. 
 But, in order to predict the rate of annihilation of dark matter particles in galactic halos, the precise size and shape of the halo is 
of paramount importance. There is still some controversy on how dark halos evolve and which shape do they have, which is reflected in the 
different parametrizations of the dark matter density around visible galaxies that are commonly used in the literature: the Navarro-Frenk-White 
(NFW) profile~\cite{Navarro:1995iw}, the Kravtsov profile~\cite{Kravtsov:1997dp}, the Moore profile~\cite{Moore:1999gc} and the 
Burkert~\cite{Burkert:1995yz} profile being the most popular ones. The common feature of these profiles is a denser spherically symmetric region 
of dark matter in the center of the galaxy, with decreasing density as the radial distance to the center increases. Where they 
diverge is in the predicted shape of the central region. Simulations of galaxy formation and evolution 
are very time consuming and complex in nature, and have not been determinant in settling the issue. Profiles obtained from N-body simulations 
tend to predict a steep power-law type behaviour of the dark matter component in the inner parts of the halo, while profiles based on observational 
data (stellar velocity fields) tend to favour a constant dark matter density near the core of the galaxies. This is the core-cusp 
problem~\cite{deBlok:2009sp}, and it is an unresolved issue which affects the signal prediction from dark matter annihilations in neutrino 
telescopes.  A general parametrization of the dark halo in galaxies can be found in~\cite{Zhao:1996mr,Yuksel:2007ac}. 
Note that the shape of the dark halo can depend on the local characteristics of any given galaxy, like the size 
of the galaxy~\cite{Ricotti:2002qu} or on its evolution history~\cite{Dekel:1986gu,Ogiya:2012jq}. 

 The shape of the dark matter halo is important because the expected annihilation signal depends on the line-of-sight \textit{(l.o.s.)}
integral from the observation point (the Earth) to the source, and involves an integration over the 
squared of the dark matter density. This is included in the so-called J-factor~\cite{Bergstrom:1997fj,Yuksel:2007ac}, 
which is galaxy-dependent, and absorbs all the assumptions on the shape of the specific halo being considered. 
In the case of our Galaxy, the expected signal from the Galactic Center using one halo 
parametrization or another can differ by orders of magnitude depending on the halo model used (see e.g. figure 1 in ~\cite{Yuksel:2007ac}).

The differential neutrino flux from dark matter annihilation from a given galaxy, $\mathrm{d}\phi_\nu / \mathrm{d}E$, depends on the candidate 
dark  matter mass, $m_{\mbox{\tiny WIMP}}$, the neutrino energy spectrum, $\mathrm{d}N_\nu / \mathrm{d}E$,  the thermally averaged product of 
the self-annihilation cross-section, $\sigma_\mathrm{A}$, times the WIMP velocity, $v$, and the  J-factor, $J_{\Psi}$,
\begin{equation}
    \frac{\mathrm{d}\phi_\nu}{\mathrm{d}E} = \frac{1}{2}\frac{\left<\sigma_\mathrm{A} v\right>}{4\pi m_{\mbox{\tiny WIMP}}^2}~ J_{\Psi} ~\frac{\mathrm{d}N_\nu}{\mathrm{d}E}
    \label{eq:fluxanni}
\end{equation}

 While a consensus on the distribution and shape of the dark halos in galaxies is achieved, neutrino telescopes usually present 
their results for a few benchmark halos. In this way the effect of different halo assumptions is factorized from other uncertainties, 
like detector systematics or the choice of the underlying particle physics model.

\begin{figure*}[t]
\centerline{
{\includegraphics[width=0.5\linewidth,height=0.4\linewidth]{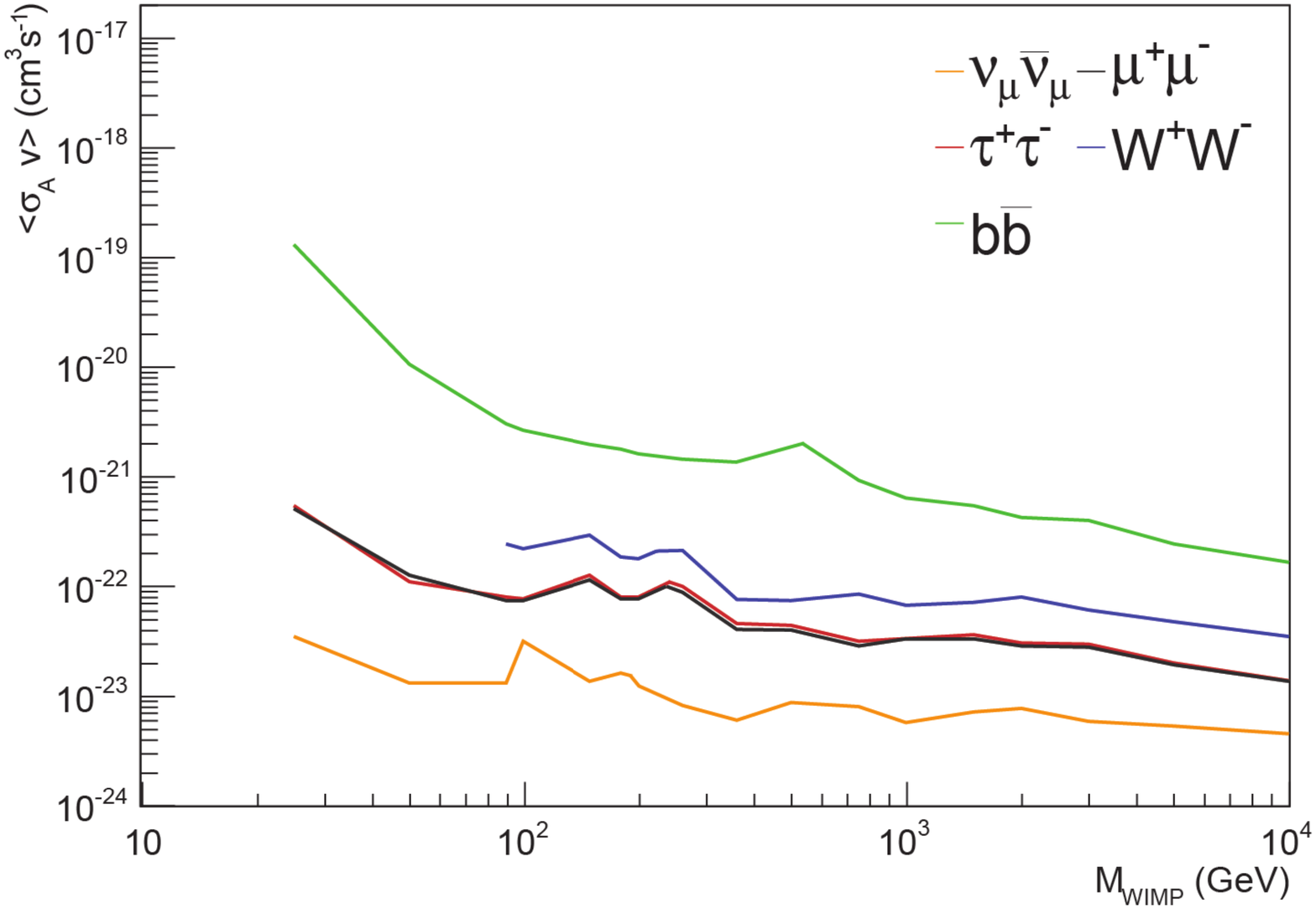}\label{fig:ANTARES_GC_limits}}
\hspace*{1pt}
{\includegraphics[width=0.6\linewidth,height=0.4\linewidth]{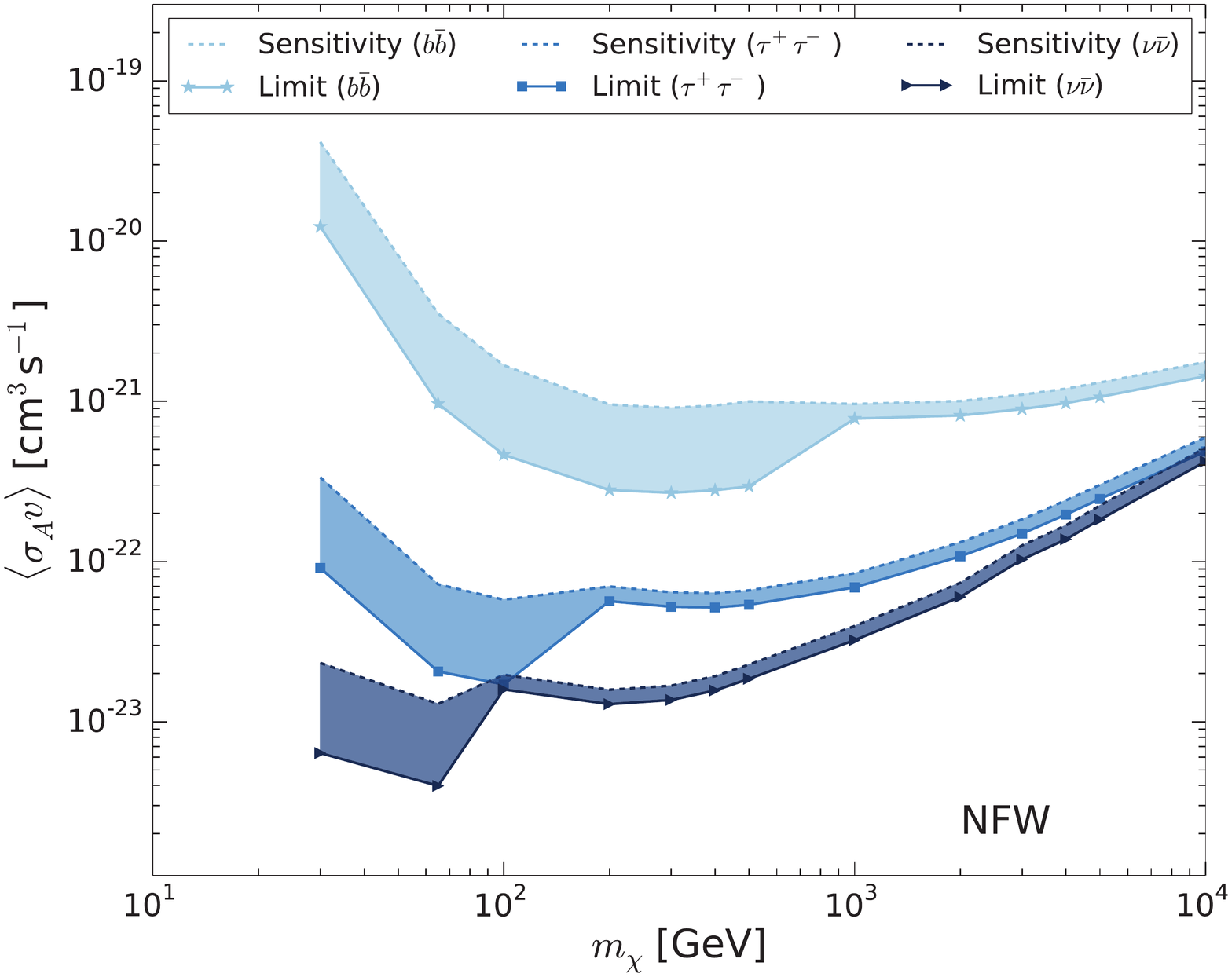}\label{fig:IC_GC_limits}}
}
\caption{
Upper limits at 90\% CL on the velocity--averaged WIMP annihilation cross section from ANTARES (left)~\cite{Adrian-Martinez:2015wey} and IceCube (right)~\cite{Aartsen:2015xej}. The limits were obtained from analyses on the Galactic Center. The different curves show different annihilation channels, assuming a 100\% branching ratio to each. The IceCube curves show the sensitivity (dashed lines) as well as the observed upper limits (solid lines). The shaded areas are to guide the eye between a sensitivity and its corresponding limit. All limits were obtained assuming an NFW-type halo profile.}
\label{fig:GCLimits}
\end{figure*}

\subsection{Observable candidates}

 The largest gravitational potential close to the Solar System is the center of our own galaxy. Further away, we find 
dwarf galaxies: small, low-brightness galaxies orbiting the center 
of the Milky Way as remnants from our Galaxy formation process. A common feature of dwarf galaxies is that they have a 
low mass-to-light ratio, suggesting the presence of large amounts of dark matter~\cite{Walker:2013a}. Since these galaxies 
are small and simple in their structure,  consisting of a small number of stars, they do not present any background from 
violent processes that could mask a signal from dark matter annihilation. The detection sensitivity 
can be further increased by stacking objects with similar characteristics. 
At cosmological scales, our closest galaxy, Andromeda, and galaxy clusters are other obvious candidates for dark matter detection. 
Andromeda is a spiral galaxy with a relatively well characterized dark matter halo~\cite{Tamm:2012hw}, while 
galaxy clusters are the largest gravitationally bound systems known and present an estimated 
85\% of dark matter in comparison with about 3\% of luminous matter, the rest consisting of intracluster gas~\cite{Voit:2004ah}.
 
\subsection{Current results}
\begin{figure*}[t]
\centerline{
{\includegraphics[width=0.5\linewidth,height=0.4\linewidth]{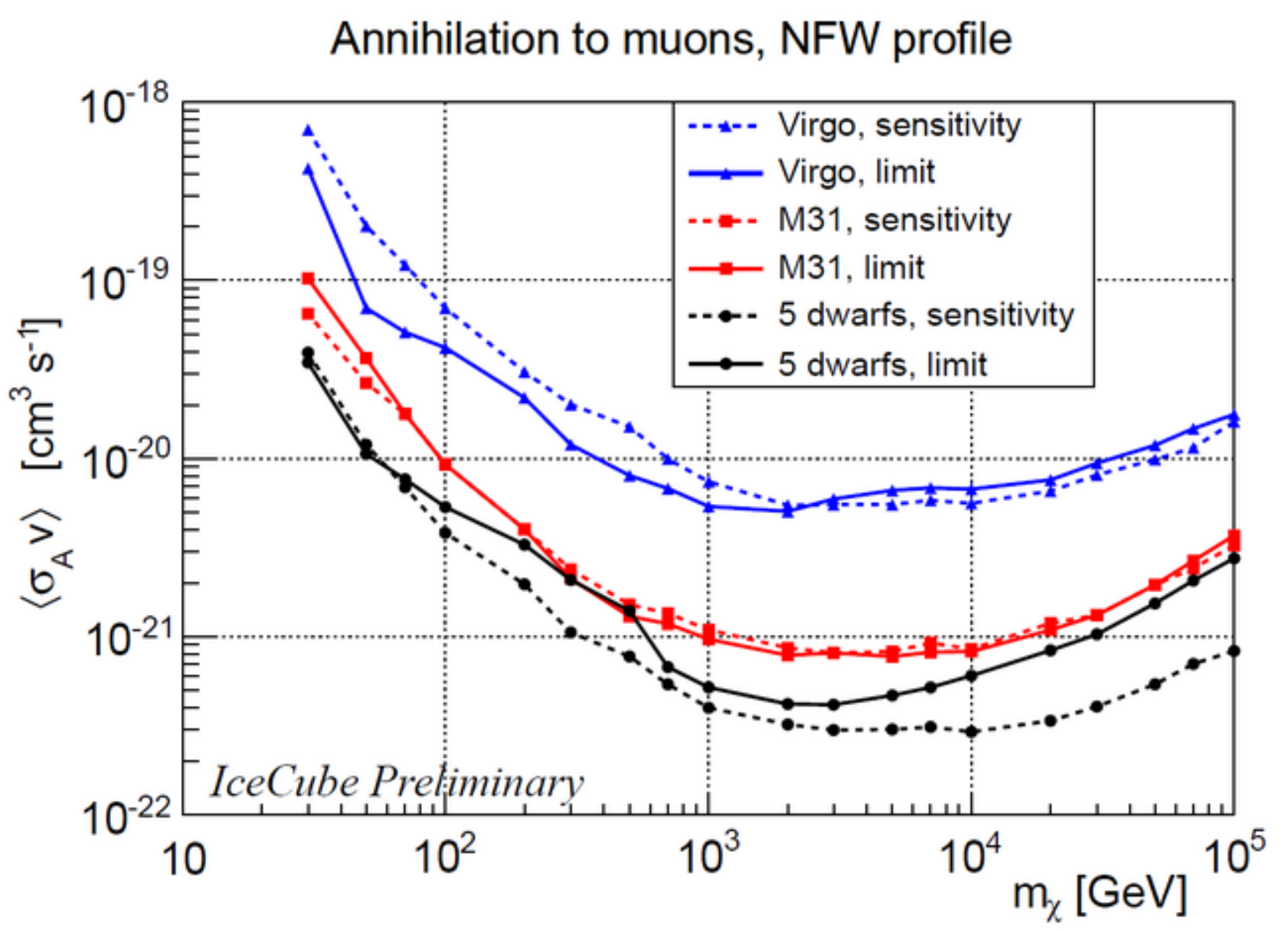}}
{\includegraphics[width=0.5\linewidth,height=0.4\linewidth]{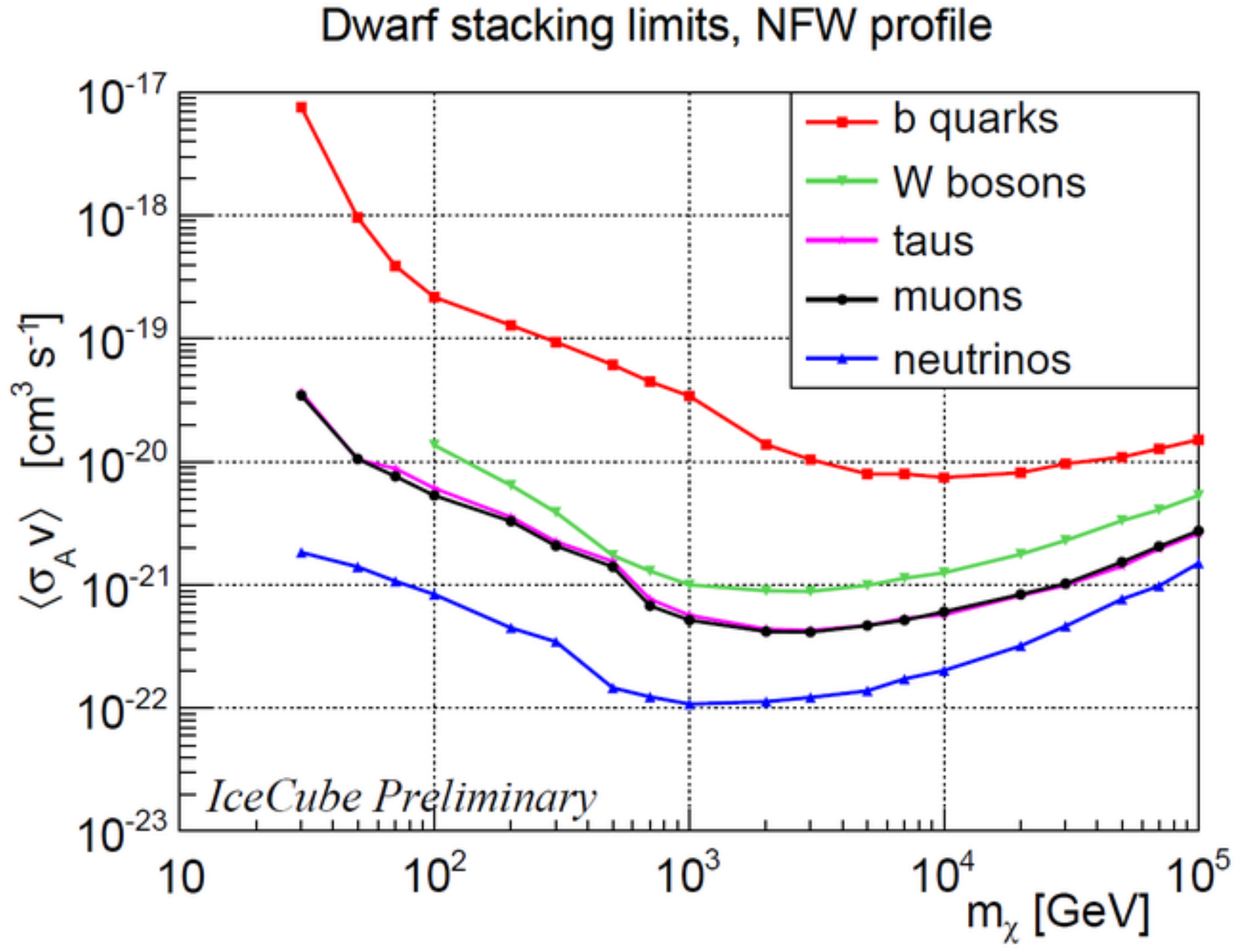}}
}
\caption{
Upper limits at 90\% CL on the velocity--averaged WIMP annihilation cross section from IceCube. 
{\bf Left:} For the Virgo cluster, the Andromeda galaxy and the stacking of five dwarf galaxies (Segue 1, Ursa Major II, Willman 1, 
Coma Berenices and Draco), assuming 100\% annihilation to muons and a NFW profile. 
{\bf Right:}  For different benchmark annihilation channels for the stacking of the mentioned five dwarf galaxies. 
Figures taken from~\cite{Aarsten:2015a}. 
}
\label{fig:dwarf_Limits}
\end{figure*}

Searches for dark matter from our own Galaxy, dwarf galaxies, Andromeda and galaxy clusters have been carried our by 
IceCube~\cite{Aartsen:2015xej,Aartsen:2013dxa,Aartsen:2014hva,Aarsten:2015a} and ANTARES~\cite{Adrian-Martinez:2015wey,ANTARES:2015a}, 
and are shown in figures~\ref{fig:GCLimits} and~\ref{fig:dwarf_Limits}. The limits obtained depend strongly on the studied galaxy, through 
the J-factor, and the assumed annihilation channel. All sources considered showed results compatible with the background-only hypothesis 
yielding limits on the velocity-averaged annihilation cross section at the level of 10$^{-20}$ cm$^3$ s$^{-1}$ -- 10$^{-23}$ cm$^3$ s$^{-1}$, 
depending on assumptions and WIMP mass.

\section{The IceCube PeV events and dark matter}\label{sec:dm_UHE}
\begin{figure*}[t]
\centerline{
{\includegraphics[width=0.5\linewidth,height=0.4\linewidth]{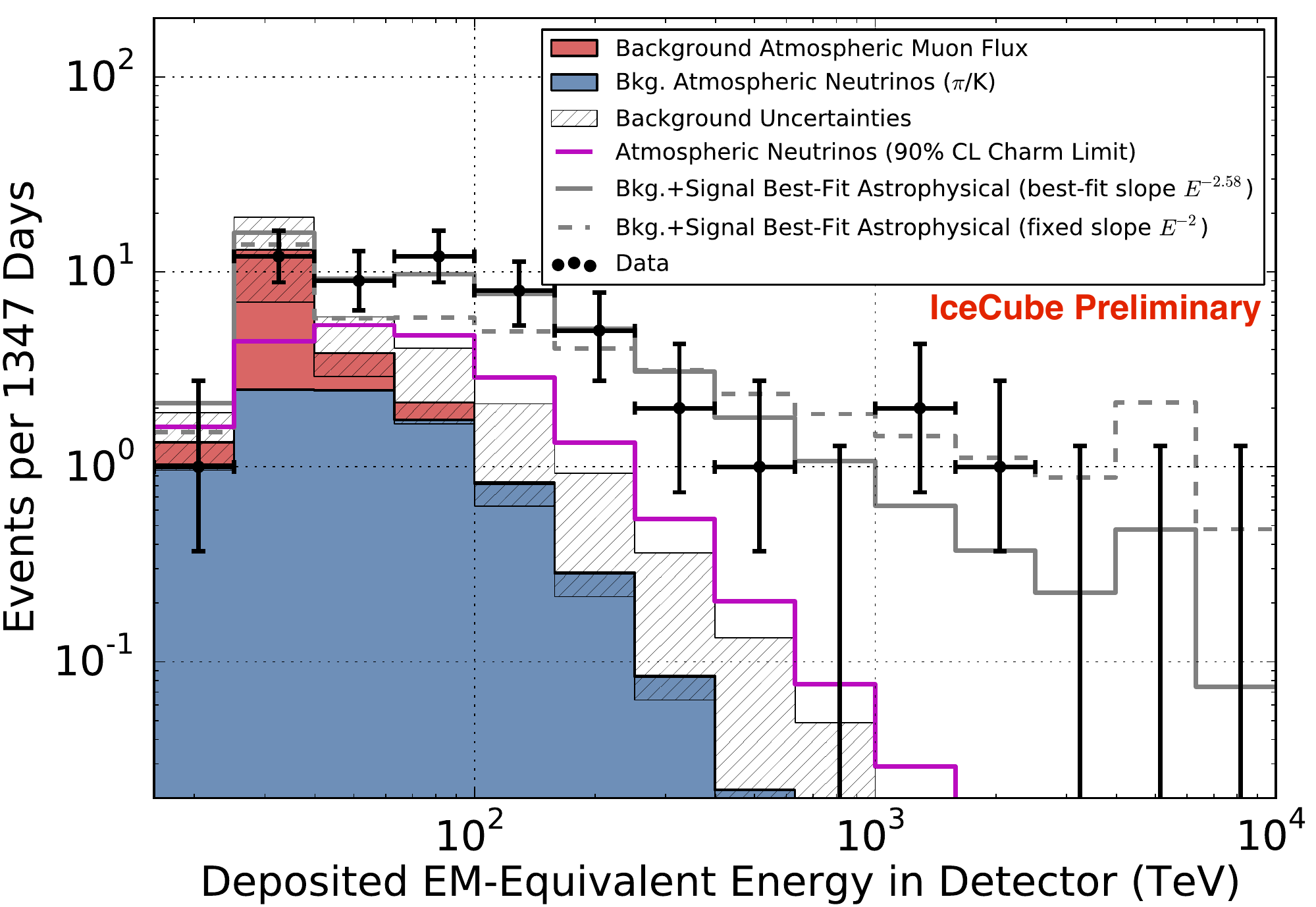}}
{\includegraphics[width=0.5\linewidth,height=0.4\linewidth]{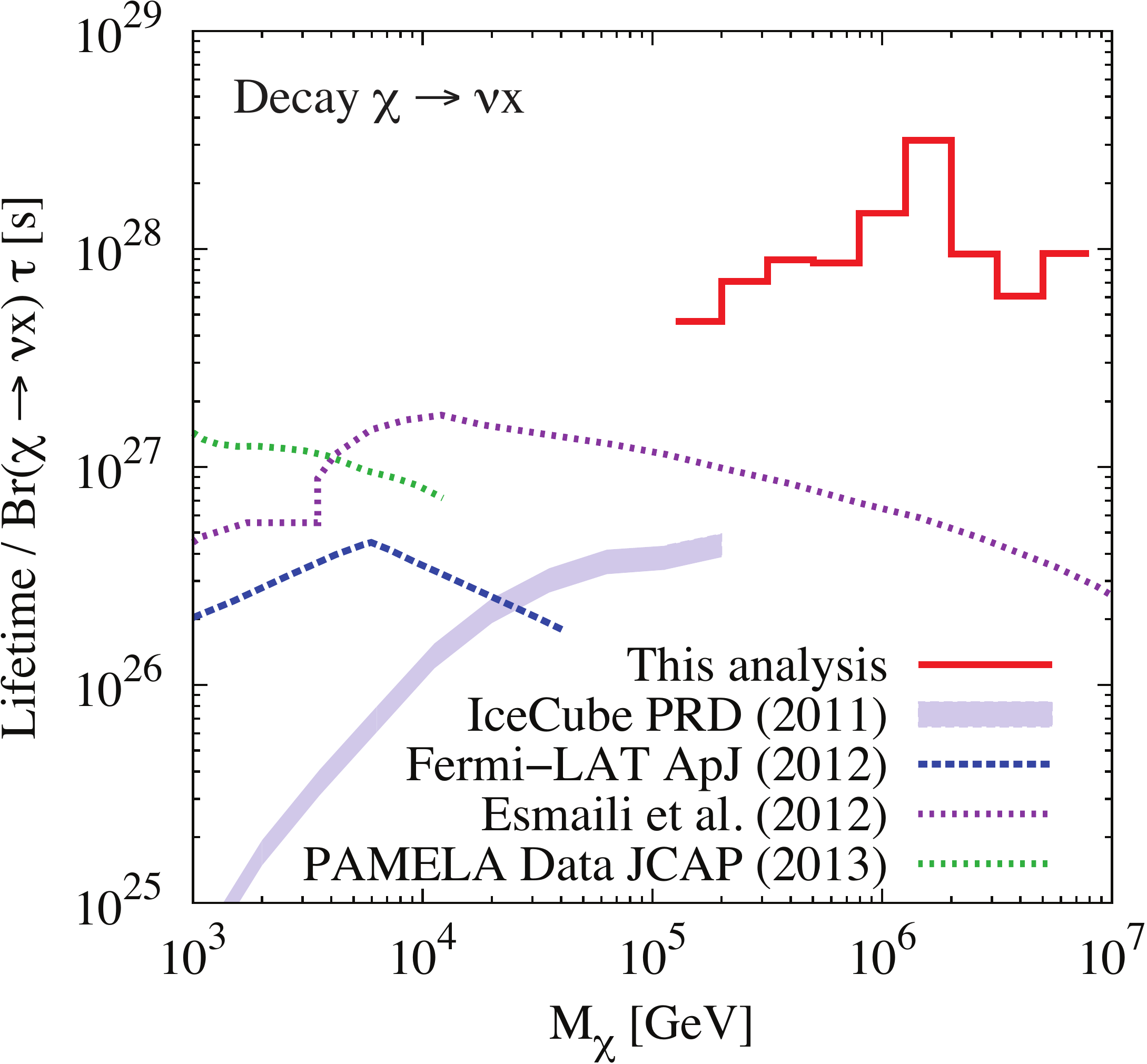}}
}
\caption{
{\bf Left:} Estimated energy deposited in the detector of the 4-year sample of ultra-high energetic IceCube events~\cite{Aarsten:2015bb}. 
{\bf Right:} Limit on the lifetime of a super-heavy dark matter candidate derived using the high-energy neutrino flux observed by IceCube (red line), 
 compared to the previous experimental constraints from IceCube~\cite{Abbasi:2011eq}, Fermi-LAT~\cite{Ackermann:2012rg}, PAMELA~\cite{Cirelli:2013hv} 
and derived limits from neutrino data~\cite{Esmaili:2012us}. Excluded are regions below the pictured lines. Figure from~\cite{Rott:2014kfa}.
}
\label{fig:X}
\end{figure*}

The recent discovery of a ultra high-energy astrophysical neutrino flux by IceCube~\cite{Aartsen:2013bka,Aartsen:2013jdh,Aartsen:2014gkd,Aarsten:2015bb} has 
triggered an intense theoretical activity trying to explain its possible origin. The still low statistics, 54 events 
detected in 1347 days of livetime with a background of atmospheric neutrinos and muons of about 21 events, and the relatively poor angular 
resolution of the cascade events, make it difficult to assign an origin to the events. The arrival directions of the events are 
compatible with an isotropic flux, maybe with a small galactic component.  Many proposed explanations are 
based on astrophysical processes, but there has been also a series of works pointing at the possible origin of the events as originating from 
heavy dark matter decay~\cite{Feldstein:2013kka,Esmaili:2013gha,Bai:2013nga,Bhattacharya:2014vwa,Rott:2014kfa,Esmaili:2014rma,Murase:2015gea,Anchordoqui:2015lqa}. 
Since the rate of dark matter decay is proportional to the dark matter density, and not to the density squared as for the annihilation case, 
the resulting neutrino flux can easily accommodate both a Galactic and a diffuse extra-galactic component of comparable strength, which is consistent 
with the distribution of the IceCube events. In the case of annihilations, a galactic component could easily dominate due to the closeness of 
our Galaxy. Another feature of dark matter decay is a monochromatic neutrino line at $m_{\mbox{\tiny WIMP}}/2$, in models with a dominant 
branching ratio to the neutrino channel. However, and more realistically, other final states will also contribute to the final neutrino spectrum 
giving a lower-energy continuum from decays into quarks and charged leptons. This is also compatible with the energy distribution of the IceCube 
events (see fig.~\ref{fig:X}) which can be 
interpreted as presenting a peak at around 2~PeV and a lower energy continuum, separated by a dip just below 1~PeV.  These are the features of 
the IceCube results that have triggered the explanations based on decaying of heavy dark matter candidates (except for the analysis 
in~\cite{Bhattacharya:2014vwa}, which concentrates on candidates in the $\cal{O}$(100)~TeV range). 
One can further argue that considering a heavy dark matter candidate is timely in view of the lack of evidence of new physics at the TeV scale 
from the LHC, which has put some strain on the vanilla WIMP paradigm with dark matter candidates on the $\cal{O}$(100)~GeV-TeV region.

 The exact nature of the potential heavy dark matter sector is however completely open in view of the IceCube data, which is not constraining 
enough at this moment. The models proposed range from ``neutrinophilic'' models where the dark matter predominantly 
decays into a $\nu\bar{\nu}$ pair, e.g.,~\cite{Anchordoqui:2015lqa,Boucenna:2015tra} to boosted dark matter scenarios, e.g.,~\cite{Kopp:2015bfa}, 
or rather model-independent analyses like in ~\cite{Bhattacharya:2014vwa} or with minimal additions to the see-saw model~\cite{Rott:2014kfa}. 
Many authors assume the existence of an astrophysical power-law diffuse component in addition to the dark matter component. 
Such combination can compensate the slight tension between the IceCube measured flux and a pure power-law assumption, although such a tension can 
be mitigated by considering a power-law with a cutoff. In any case, the results from the different analyses of the IceCube data tend to concur 
on a limit on the lifetime of a generic dark matter candidate at the level of $\geqslant \cal{O}$(10$^{27}$)~s. 

Decaying dark matter has also been proposed as an explanation to the observed $e^+$ excess in cosmic rays~\cite{Adriani:2008zr,Chang:2008aa,FermiLAT:2011ab,Aguilar:2013qda}, 
although a slight fine tuning of the models towards leptophilic dark matter candidates (decaying preferably to charged leptons rather than to quarks) 
seems necessary, since new-physics in the proton spectrum is strongly constrained by data from the same detectors~\cite{Adriani:2011cu,Consolandi:2014uia}. 
The IceCube data can provide an escape route by confirming a neutrino annihilation channel while complying with current limits on annihilation 
to charged leptons from cosmic-ray detectors. But we must wait for more events in IceCube

\section{Outlook}\label{sec:dm_outlook}
 Naturally, any model of dark matter producing a neutrino flux must be viewed in the grand scheme of dark matter searches and be consistent 
with limits from the $\gamma$ and cosmic-ray channels~\cite{Ackermann:2015zua,Aleksic:2013xea,Grube:2012fv}, as well as from constrains from 
direct~\cite{Aalseth:2011wp,Behnke:2012ys,Aprile:2013doa,Akerib:2013tjd} and accelerator searches~\cite{Cakir:2015gya}. The near future 
will bring us more events from the neutrino telescopes in operation, as well as the next-generation, large-mass direct search 
experiments~\cite{Aprile:2012zx,Baudis:2012bc} and the directional detection efforts~\cite{Gehman:2013mra,Battat:2014van,Alexandrov:2014gda}, which 
will provide a quantitative jump in the dark matter search paradigm. Neutrino telescopes 
are also planning the next generation arrays, PINGU~\cite{Aartsen:2014oha} and ORCA~\cite{Katz:2014tta} on the low-energy side, and 
the high-energy extension of IceCube~\cite{Aartsen:2014njl}, KM3NET~\cite{Coniglione:2015aqa} and GVD~\cite{Avrorin:2013uyc} at the high energy end. 
There is both overlap and complementarity in the WIMP mass range and annihilation channels covered by all these experiments. 
It is just left to Nature to reveal what solution she has chosen as dark matter.\\

\noindent{\bf Acknowledgments:} I am indebted to M. Rameez, J. Zornoza, C. T\"onnis and S. Demidov for kindly providing some of the data shown in figure~\ref{fig:SunLimits}.


\end{document}